\documentclass[12pt]{article}
\usepackage{graphicx}% Include figure files
\usepackage{amssymb,amsmath}

\newcommand{\beq}{\begin{equation}}
\newcommand{\eeq}{\end{equation}}
\newcommand{\bea}{\begin{eqnarray}}
\newcommand{\eea}{\end{eqnarray}}
\newcommand{\citep}[1]{\cite{#1}}
\newcommand{\cbr}{}
\author{Robert W. Johnson \\%\thanks{Thanks to the editors of this wonderful journal!}\\
\small Alphawave Research\\[-0.8ex]
\small Atlanta, GA, USA\\
\small \texttt{robjohnson@alphawaveresearch.com}\\}
\title{Edge Adapted Wavelets, Solar Magnetic Activity, and Climate Change}
\date{\today}
\begin{document}
\maketitle

%% Abstract
\begin{abstract}
The continuous wavelet transform is adapted to account for signal truncation through renormalization and by modifying the shape of the analyzing window.  Comparison is made of the instant and integrated wavelet power with previous algorithms.  The edge adapted and renormalized admissible wavelet transforms are used to estimate the level of solar magnetic activity from the sunspot record.  The solar activity is compared to Oerlemans' temperature reconstruction and to the Central England Temperature record.  A correlation is seen for years between 1610 and 1990, followed by a strong deviation as the recently observed temperature increases.
\end{abstract}

%% Keywords \keywords{wavelet; sunspot; solar activity; climate change}

%  BEGIN ARTICLE TEXT
%\linenumbers*

\section{Introduction}
The continuous wavelet transform has become a popular tool in the analysis of data signals~\citep{Morlet:1984,kaiser:1994,Frick:1997670F,torrence:98,fligge-313,polygian-725,ligao-181}, particularly among those investigating variations in global climate~\citep{piscaron-1661,weber-917,echera-41}.  Essentially just a special type of windowed Fourier transform, its primary drawback is the loss of transform response attributed to the finite length of a truncated signal, known as the cone-of-influence.  Various algorithms have been proposed to correct for its effect~\citep{Foster:1996,Frick:1997426,sweldens98lifting,rwj-jgr02}; however, none have proven entirely satisfactory.  In this article, we investigate the proper renormalization for a truncated wavelet as well as propose an algorithm which modifies the shape of the analyzing window with respect to both the location of the wavelet and the length of the signal.  These algorithms produce a reasonable estimate for the integrated wavelet power even as they approach the edges of the known data, verified by extensive evaluation of test signals.

Armed with these transforms, we calculate the level of solar magnetic activity as given by the sunspot record using both the international~\citep{sidc:online} and group~\citep{hoytschatten:98} sunspot numbers, consistent with our previous estimate~\citep{rwj-jgr02}.  Our indicators of climate change are the global temperature reconstruction by~\cite{Oerlemans:2005} and the Central England Temperature record~\citep{hadcet:92}.  Plotting the temperature against the solar magnetic activity reveals a correlation which extends up until about 1990, after which a strong deviation from the linear trend is noted as the recently observed temperature increases despite a decreasing level of solar activity.  The most recent year for which data is available reapproaches the trend.

\section{Normalization, Central Frequency, and Admissibility}
\begin{figure}[t]%[tb]
\includegraphics[width=\columnwidth]{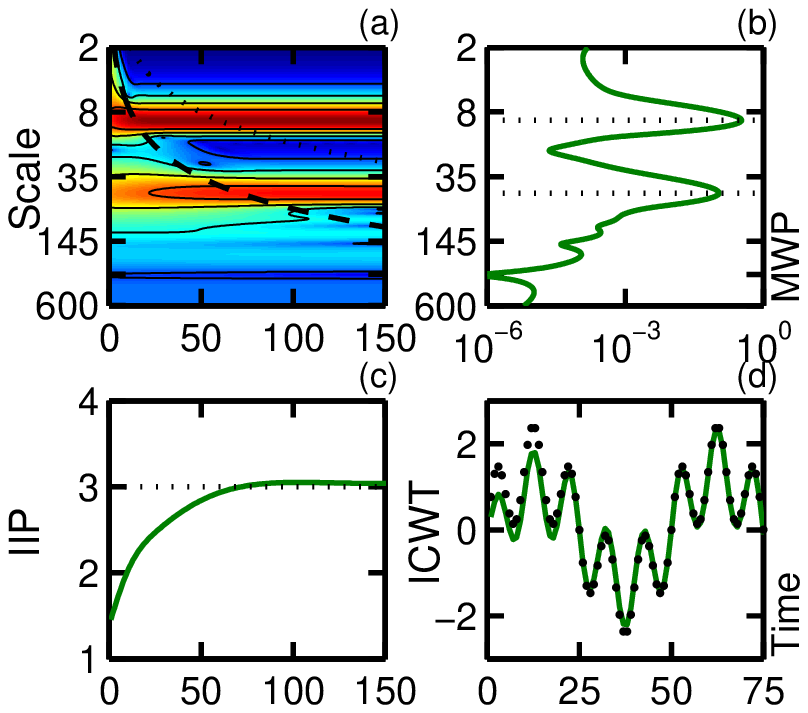}
\caption{CWT analysis of a signal with elements of amplitude 1 and $\sqrt{2}$ and period 10 and 50. (a) The power spectral density with the cone-of-influence (dashed) and cone-of-admissibility (dotted) overlaid.  An unlabelled tick on the scale axis appears at the signal duration $N_t = 300$. (b) Mean wavelet power displaying the signal peaks as well as the large scale response to the signal truncation.  A trough appears at the scale of the signal length. (c) The integrated instant power decreases to half its expected value towards the signal edge.  (d) The reconstruction is diminished compared to the signal (dots) near the edge} %% no full stop at the end of caption
\label{fig:A}
\end{figure}
We begin by writing the complex Morlet wavelet $\psi^0_{s,t}(t')$ in its conventional normalization at scale $s = 1 / f_s = 2 \pi / \omega_s$ and offset $t$ with unity sample rate $\Delta t' \equiv 1$ as the product of a scale dependent constant $C$, a normalized window $\Phi$, and a normalized wave $\Theta$ using the parameter $\eta \equiv (t' - t) / s$, \bea
\psi^0_{s,t}(t') \equiv C^0_s \Phi^0_{s,t}(t') \Theta^0_{s,t}(t') = \pi^{-1/4} s^{-1/2} e^{- \eta^2/2} e^{i \omega_1 \eta} \;,
\eea where $\omega_1 \approx 2 \pi$ is the central frequency of the mother wavelet at unity scale and zero offset, $\psi^0_{1,0}(t') = \pi^{-1/4} e^{- {t'}^2/2} e^{i \omega_1 t'}$. In practice, {\cbr the window has a finite extent of $- \lfloor s \chi \leq t' \leq \lfloor s \chi$ and the wavelet a length of $ N_{t'} = 2 \lfloor s \chi + 1$, where the parameter $\chi$ defines the resolution width beyond which the Gaussian is taken to be negligible; here we use $\chi = 6$}.  The mother wavelet is normalized to unit energy $\int_{-\infty}^\infty \vert \widetilde{\psi}^0_{1,0}(\omega) \vert^2 d\omega = 1$ so that the Fourier transform of $\psi^0(t'/s)$ is $\widetilde{\psi}^0(s \omega) = \sqrt{2 \pi s} \, \widetilde{\psi}^0_{1,0}(s \omega)$.  We have found previously that the inclusion of additional factors in the scale dependent constant $C_s = \sqrt{2} \, C^0_s / s$ produces a transform with some very desirable properties, \bea
\psi_{s,t}(t') = \sqrt{2} \, \pi^{-1/4} s^{-3/2} e^{- \eta^2/2} e^{i \omega_1 \eta} \;.
\eea  The factor $\sqrt{2}$ comes from absorbing the transform response at negative scales, usually neglected, so that the mother wavelet now has a norm of 2, and the factor $s^{-3/2}$ arises from symmetrization of the forward and inverse wavelet transforms of a signal $y(t) = Re (A_y e^{i \omega_y t})$ with duration $N_t$ by pulling over a factor $1/s$ usually left in the denominator of the integrating measure ({\it cf.} Equations (6) and (9) by~\cite{Frick:1997426}), \bea 
&{\rm CWT}(s,t) =& \sum_{t'} \psi_{s,t}^*(t') y(t') \;,\\
&{\rm ICWT}(t) =& Re \left[\sum_s \sum_{t'} \psi_{s,t}^*(t') {\rm CWT}(s,t') \Delta s \right] \;,
\eea noting that $\psi^*(\eta) = \psi(-\eta)$, so that the scaled wavelet now has a norm of $2/s^2$.  This normalization produces a power spectral density ${\rm PSD}(s,t) \equiv \vert \sqrt{2} \, {\rm CWT} \vert^2$ such that the integrated area of {\cbr an isolated} peak in the instant wavelet power ${\rm IWP}_t(s) \equiv {\rm PSD}(s,t)$ returns the square of the amplitude $A_y$ of the signal element.  Noting that the window has imparted the particle-like property of locality, we appeal to the photon whose energy is proportional to its frequency $E_\nu \propto \nu = 1 / s_\nu$ yet must extend for at least one cycle period to write its power as $\propto 1/ s_\nu^2$, {\cbr suggesting a physical basis for our choice of normalization}.  The factor $\sqrt{2}$ applied to the amplitude of the CWT when forming the PSD we attribute to a root-mean-square effect, such that if we asked for the rms PSD, no factor would be necessary here but would then be introduced to compare the estimate for the rms amplitude to the peak amplitude $A_y$.  The margins of the PSD yield the mean wavelet power ${\rm MWP}(s) = N_t^{-1} \sum_t {\rm PSD}(s,t)$ and the integrated instant power ${\rm IIP}(t) = \sum_s {\rm PSD}(s,t) \Delta s$.  The analysis of a test signal of duration 300 time units with signal elements of amplitudes 1 and $\sqrt{2}$ and periods 10 and 50 respectively is shown in Figure~\ref{fig:A}.  The cone-of-influence is defined as the $e$-folding time, and the more restrictive cone-of-admissibility indicates where wavelets are wholly contained by the extent of the signal.  Apparent is the loss of transform response and reconstruction as the signal edge is approached, and the slight overestimation of the power is attributed to the response at large scales to the signal truncation.
\begin{figure}[t]%[tb]
\includegraphics[width=\columnwidth]{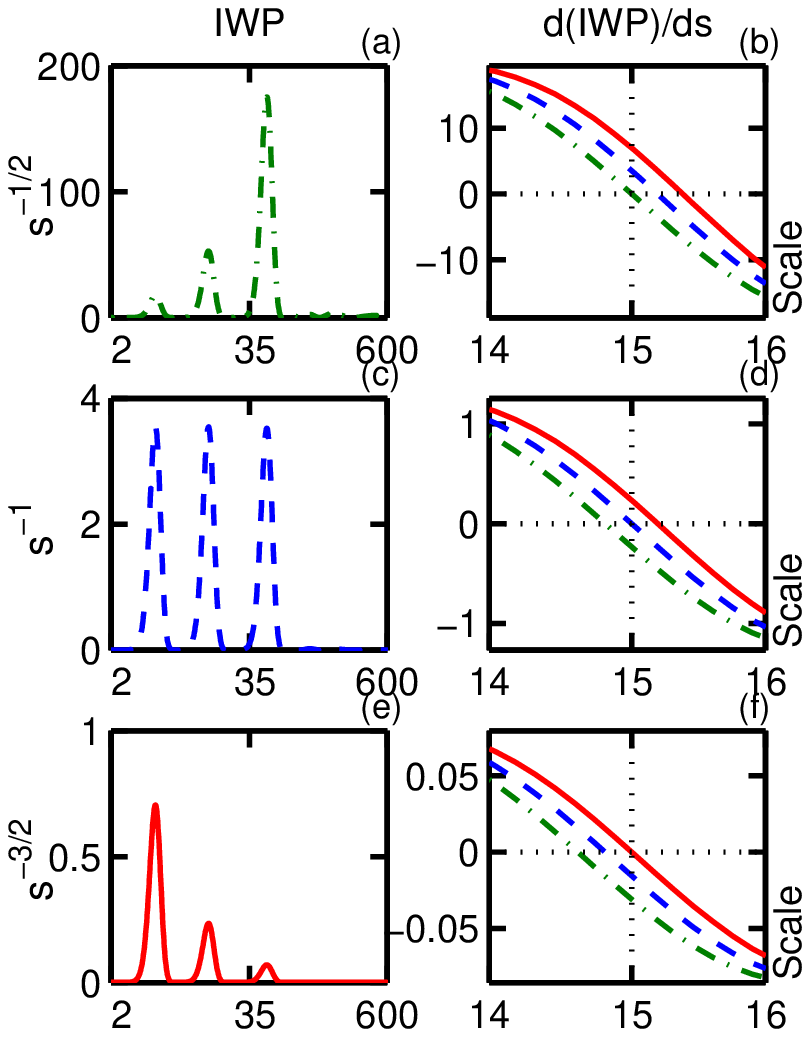}
\caption{CWT analysis of a signal with elements of unit amplitude and periods 5, 15, and 50 for central frequencies $2 \pi - 1 / 4 \pi$ (dash-dot), $2 \pi$ (dashed), and $2 \pi + 1 / 4 \pi$ (solid) and forward transform scalings of $s^{-1/2}$, $s^{-1}$, and $s^{-3/2}$ labelled by the vertical axis.  The left column is the instant wavelet power at $N_t/2$ and the right is its gradient {\cbr for all three central frequencies in the vicinity of the signal peak at period 15}. The area under each peak in (e) is unity} %% no full stop at the end of caption
\label{fig:B}
\end{figure}
\begin{figure}[t]%[tb]
\includegraphics[width=.75\columnwidth]{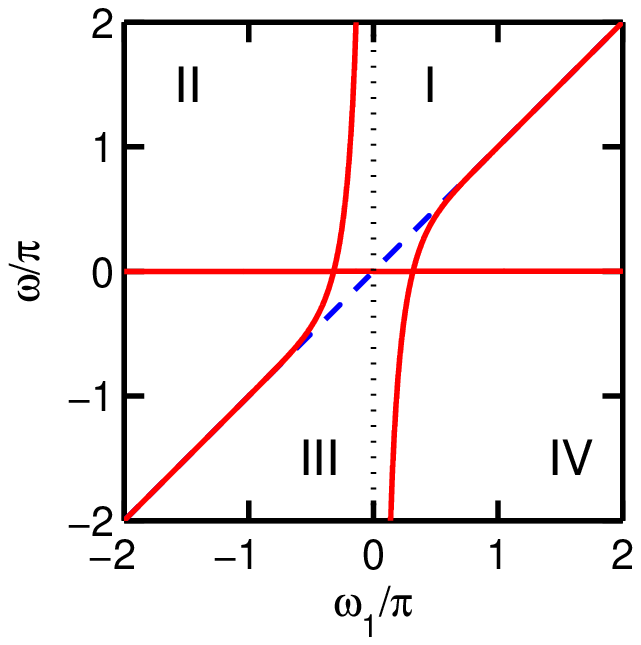}
\caption{The solution (solid) of the equation $\omega_1 e^{- \omega_1 \omega} = \omega_1 - \omega$ derived by setting $d \widetilde{\psi}_1(\omega) / d \omega_1 = 0$  crosses the $\omega_1$ axis at $\pm 1$. The inadmissible solution (dashed) is linear} %% no full stop at the end of caption
\label{fig:C}
\end{figure}

The central frequency given by~\cite{torrence:98} to unify the Fourier period and wavelet scale, $\lambda_1 / s_1 = 4 \pi / (\omega_1 + \sqrt{2 + \omega_1^2}) = 1$ yielding $\omega_1 = 2 \pi - 1 / 4 \pi$, is no longer appropriate for our normalization.  Using a test signal of the same duration with elements of unit amplitude and periods 5, 15, and 50, we consider transforms with central frequencies $2 \pi - 1 / 4 \pi$, $2 \pi$, and $2 \pi + 1 / 4 \pi$ and forward transform scalings of $s^{-1/2}$, $s^{-1}$, and $s^{-3/2}$ appearing in the CWT.  The left column in Figure~\ref{fig:B} displays the instant wavelet power at the center of the transform $N_t/2$ {\cbr for a single central frequency (the others are shifted by a small amount), and the right column shows its gradient for all three central frequencies in the vicinity of the signal peak at period 15}; similar graphs obtain for the other peaks, noting that the forward scaling of $s^{-1}$ in Figures~\ref{fig:B}(c) and (d) corresponds to that recently proposed by~\cite{liu-2093}.  Only the transform with scaling $s^{-3/2}$ produces peaks with an integrated area equal to the sum of the squares of the amplitudes, and we note that the locations of its peaks coincide with the signal periods for the central frequency of $\omega_1 = 2 \pi + 1 / 4 \pi$.

In order to satisfy the admissibility condition $\widetilde{\psi}_1(0) = 0$, one must use the zero mean formula {\cbr so that the wavelet has no zero frequency response.  Neglecting normalization and offset, the Fourier transform} \bea {\cbr
\widetilde{\psi}^0_1(\omega) = \int_{-\infty}^\infty \psi^0_1(t') e^{-i \omega t'} dt' \propto e^{-(\omega_1 - \omega)^2/2} }
\eea indicates the mother wavelet contains a DC component $\widetilde{\psi}^0_1(0) \propto e^{- \omega_1^2/2} \equiv d_c \sim 10^{-9}$.  From $\Theta^0 \equiv e^{i \omega_1 \eta}$ one must subtract the continuous admissibility term, so that the zero mean wave becomes $\Theta = \Theta^0 - d_c$, defining the admissible wavelet transform AWT. {\cbr  With this subtraction, the Fourier transform of the admissible mother wavelet acquires a term which cancels the DC component \bea
\widetilde{\psi}_1(\omega) &=& \int_{-\infty}^\infty C_s \Phi^0_1(t') \Theta_1(t') e^{-i \omega t'} dt' \;, \\
&\propto& e^{-(\omega_1 - \omega)^2/2} - e^{-(\omega_1^2 + \omega^2)/2} \;,
\eea and setting its derivative with respect to $\omega_1$ \bea
\dfrac{d \widetilde{\psi}_1(\omega)}{d \omega_1} \propto (\omega-\omega_1) e^{-(\omega_1-\omega)^2/2} + \omega_1 e^{-(\omega_1^2+ \omega^2)/2}
\eea equal to 0 to reveal the extrema yields the equation \bea
\omega_1 - \omega = \omega_1 e^{- \omega_1 \omega} \;,
\eea where the effect of the zero mean formula $d_c$ is carried through on the right hand side}.  Its solution is given graphically in Figure~\ref{fig:C} alongside $\omega_1 = \omega$, and its interpretation in the first and third quadrants {\cbr as the wavelet response to positive and negative frequencies} is straightforward, less so for the second and fourth {\cbr where the second derivative indicates a minimal response}.  The solution crosses the $\omega_1$ axis at $\pm 1$, and the effect of the admissibility term is to induce a nonlinear response as the signal frequency approaches zero.

Details of the edge adaption have been relegated to the Appendix.  Our previous algorithm, identified as the corrected continuous wavelet transform CCWT, is found to overestimate the power carried by low frequency components.  The newly proposed edge adapted EAWT and renormalized admissible RAWT transforms perform much better at estimating the power carried by the signal.  This performance comes at a cost to the reconstruction for the EAWT, as a trade-off between frequency response and power estimation {\cbr has been} made.

\begin{figure}[t]%[tb]
\includegraphics[width=\columnwidth]{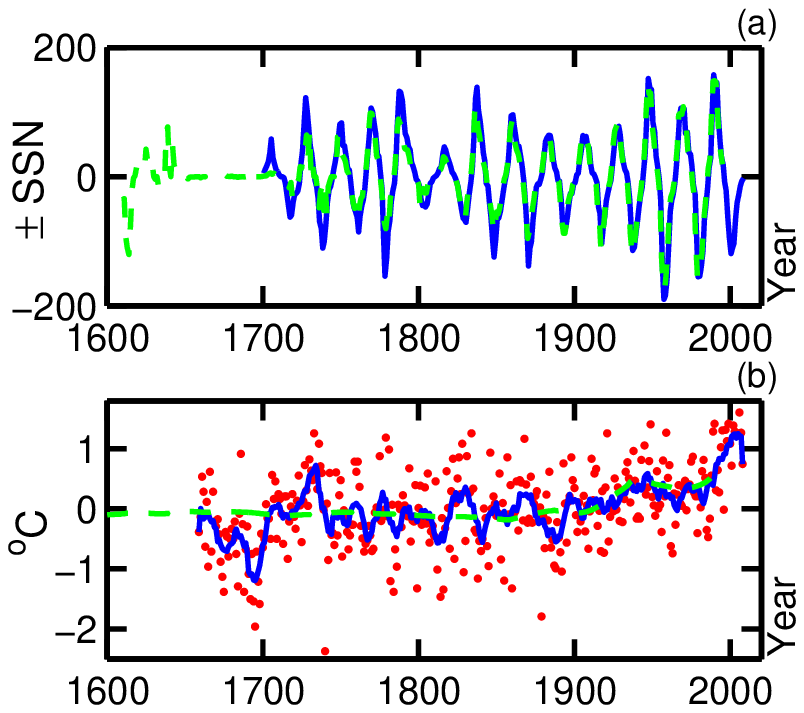}
\caption{(a) Derectified signal for the group sunspot number $R_g$ (dashed) and the international sunspot number $R_i$ (solid).  (b) Mean subtracted glacial temperature reconstruction GTR (dashed) along with the Central England Temperature CET (dots) and 9-year smoothed CET (solid)} %% no full stop at the end of caption
\label{fig:I}
\end{figure}

\section{Data Selection}
Turning now to the data chosen for this analysis, our indicator of solar magnetic activity will be the integrated instant power of the yearly sunspot number.  As it is the underlying magnetic activity characterized by the Hale cycle of $\sim$22 years in which we are interested, we must derectify the signal given by the sunspot number; doing so has the advantage of keeping the cycle shapes unaltered compared to simply analyzing the mean-subtracted rectified signal.  Previously, we followed~\cite{BuckandMac:1992,BuckandMac:1993} in taking the square root of the sunspot number and alternating the sign as appropriate, but after a thorough evaluation of the arguments presented by~\cite{bracewell-53,bracewell-88}, we now only alternate the sign to let any nonlinearity show up naturally as harmonics in the spectrum and subtract the mean before entering the wavelet analysis.  We will consider both the sunspot group number $R_g$ by~\cite{hoytschatten:98}, which covers the years 1610 through 1995, as well as the international sunspot number $R_i$ by the~\cite{sidc:online}, which continues the Wolf index from 1700 up until the present day.  While $R_g$ starts earlier and displays a more pronounced secular trend~\citep{wilson-88}, $R_i$ is known to display a tighter correlation with sunspot area and radio flux~\citep{hathaway-357}; their relative utility depends on whether one is more interested in the past or the future.  The derectified signals are shown in Figure~\ref{fig:I}(a).

Our indicators of climate change are the glacial reconstruction of global temperature by~\cite{Oerlemans:2005} and the Central England Temperature record~\citep{hadcet:92}, ``the longest instrumental record of temperature in the world.''\footnote{Met Office Hadley Centre for Climate Change}  The glacial temperature reconstruction GTR spans the years 1600 through 1990 and combines data from around the globe, while the CET data begin in 1659 and continue to the present for a specific locality, both displayed in Figure~\ref{fig:I}(b) with their mean subtracted.  Also shown is the 9-point running average of the CET, which generally oscillates about the GTR.  The recent warming trend is apparent in both data sets, {\cbr and the latest data indicate a downturn in observed temperature}.  Note that the GTR is defined as the ``anomaly'' and has had a constant subtracted by its originator.

\section{Analysis of the Data}
\begin{figure}[t]%[tb]
\includegraphics[width=\columnwidth]{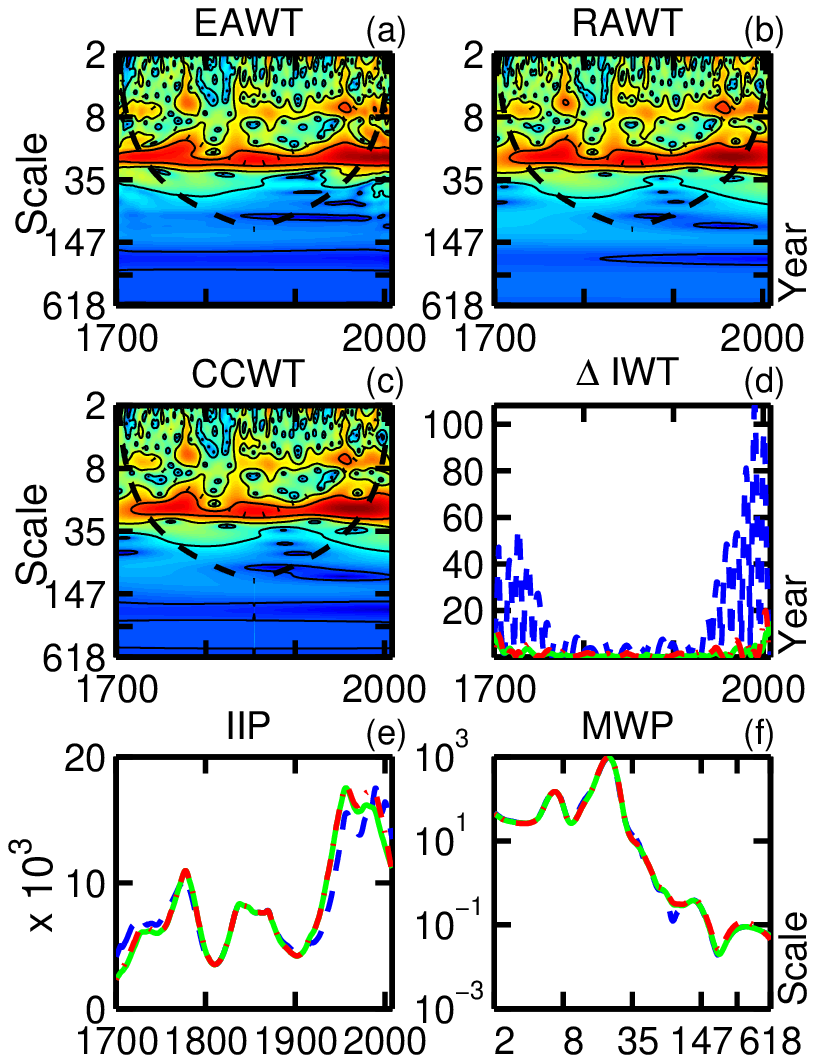}
\caption{International sunspot number $R_i$ power spectral density (a)-(c), quality of reconstruction (d), integrated instant power (e), and mean wavelet power (f) for the EAWT (dashed), the RAWT (solid), and the CCWT (dash-dot).  $\Delta\, {\rm IWT}$ is the absolute value of the discrepancy between the reconstruction and the signal} %% no full stop at the end of caption
\label{fig:J}
\end{figure}

\begin{figure}[t]%[tb]
\includegraphics[width=\columnwidth]{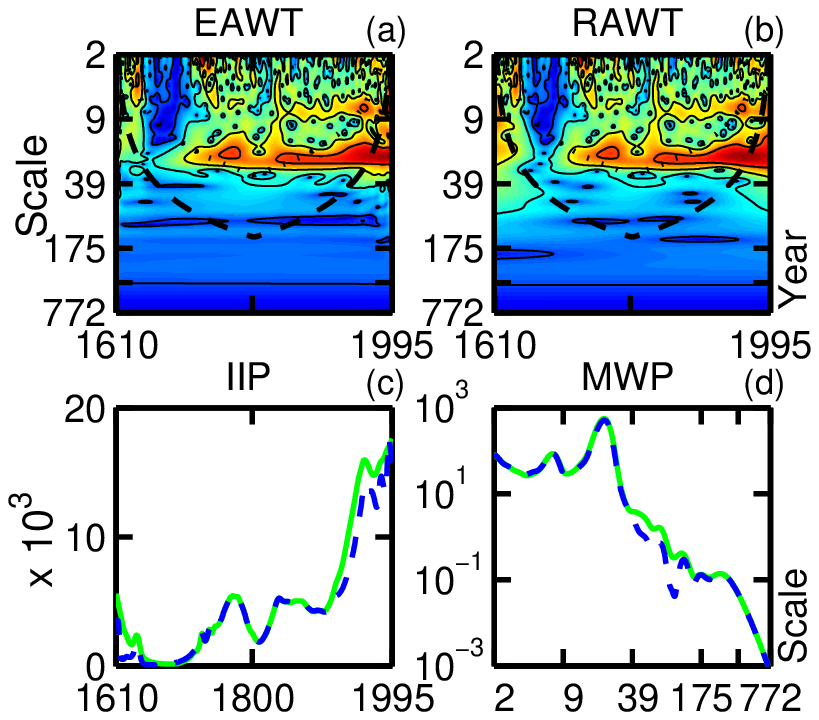}
\caption{Group sunspot number $R_g$ power spectral density (a)-(b), integrated instant power (c), and mean wavelet power (d) for the EAWT (dashed), the RAWT (solid)} %% no full stop at the end of caption
\label{fig:K}
\end{figure}

Beginning with the signal for $R_i$ taken up until 2008, we display its EAWT, RAWT, and CCWT in Figure~\ref{fig:J}.  The quality of reconstruction is not as good for the EAWT, as measured by $\Delta\, {\rm IWT}$ the absolute value of the discrepancy with the signal, which is several factors larger than for the other transforms.  The IIP for the CCWT generally agrees with that for the RAWT yet exceeds it for recent years, and the EAWT has a slightly different behavior for decades at either edge of the data.  This total power estimate compares well with that found by~\cite{cjaa:46578} determined from just the Schwabe cycle.  The spectra given by the MWP are in close agreement, except for a region around the 100 year scale where the response of the EAWT has dropped, with the third harmonic at $\sim$7 years clearly present.  The trough marking the extremely low frequency ELF region is not located precisely at the scale of the duration.  Discarding the CCWT, the results for $R_g$ are shown in Figure~\ref{fig:K}.  The IIPs of the RAWT  and the EAWT again display some variation for recent decades.  The MWP for $R_g$ does not display as prominent a third harmonic, and again the response of the EAWT is slightly diminished for scales between the maximum scales of admissibility and influence.  No trough appears at the duration scale, indicating significant ELF contribution.  We take the average of the IIPs for the EAWT and the RAWT integrated to the duration scale as our indicator of solar magnetic activity SMA for both $R_i$ and $R_g$; the ELF contribution is generally less than 1 percent of the total.

\begin{figure}[t]%[tb]
\includegraphics[width=\columnwidth]{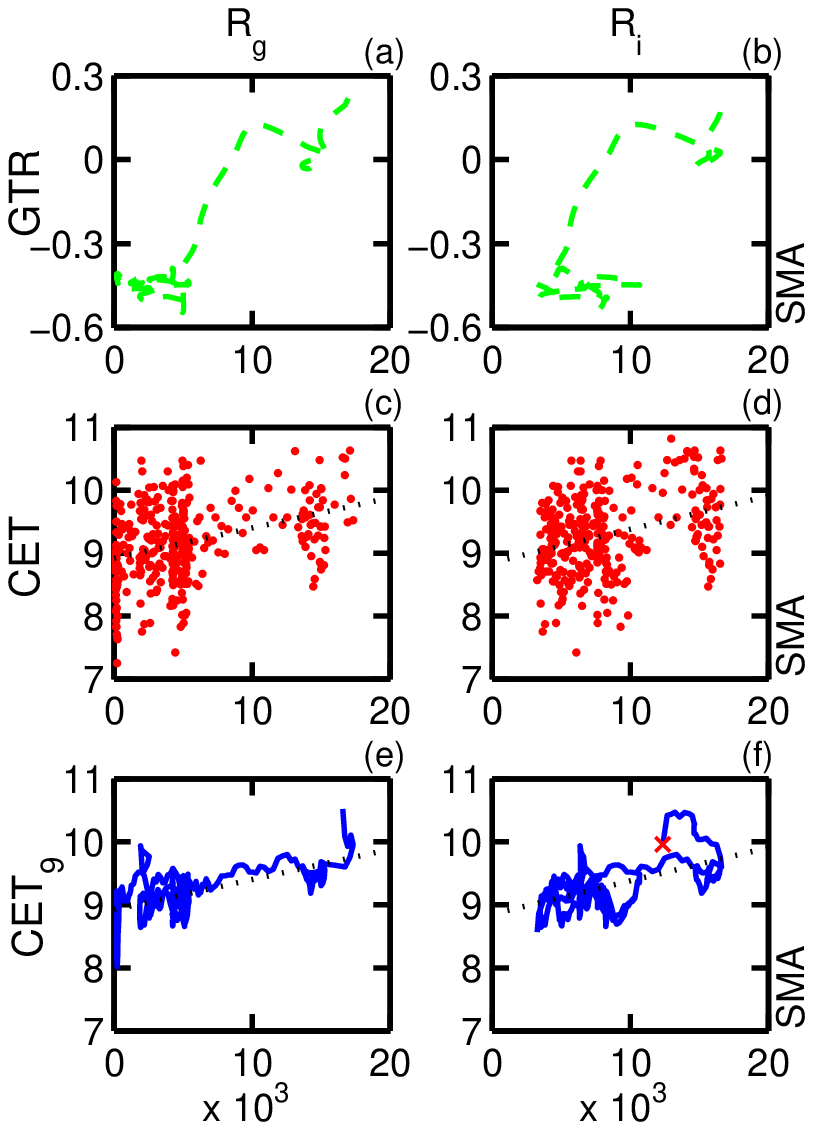}
\caption{ {\cbr Scatter-plot of the glacial temperature reconstruction (a)-(b) and the Central England Temperature record (c)-(f) {\it versus} solar magnetic activity for the group number (left column) and the international number (right column).  Shown is their projection along the time axis.  The dotted line on the CET plots displays a linear fit, and ${\rm CET}_9$ uses the 9 point running average to reduce the scatter.  The data for $R_g$ end in 1995, and the point for year 2008 is marked with an $\times$ in (f)} } %% no full stop at the end of caption
\label{fig:L}
\end{figure}

Finally we are in position to compare the level of solar magnetic activity with our indicators of climate change. {\cbr  To produce our comparison, we plot the time series of temperature in Figure~\ref{fig:I}(b) against the time series of solar magnetic activity from Figures~\ref{fig:J}(e) and~\ref{fig:K}(c) and project along the time axis.}  These results are similar to our previous work~\citep{rwj-jgr02}, but we feel that here we have a better estimate of the SMA; the similarity indicates that the result is robust.  In Figures~\ref{fig:L}(a) and (b) we plot the global temperature reconstruction GTR against the SMA for $R_g$ and $R_i$ respectively.  No relation is seen until the warming trend of the 20th century begins, then both the GTR and SMA increase to their highest levels.  We learn from~\cite{Oerlemans:2005} that measurements for the GTR before 1900 becomes increasingly scarce, thus the lack of a visible relationship for the first three centuries may result from an inadequate supply of data.  When the CET is plotted against the SMA in Figures~\ref{fig:L}(c) and (d), we find a band emerges within which the points are scattered.  A linear fit to the points is shown by the solid line, and the quality statistic $Q \equiv {\rm erfc} (\vert r_P \vert \sqrt{N/2})$ determined by Pearson's $r$ as well as the baseline temperature $T_b$ at 0 SMA are displayed in Table~\ref{tab:A} and compared with values obtained when applying the square root to the sunspot number as previously.  Interestingly, the warming attributed to the increase in SMA by $R_g$ is .83 degrees, nearly the same as the total warming in the GTR of .78 degrees.  The relation becomes more apparent when one uses the 9-point running average of the CET, Figures~\ref{fig:L}(e) and (f).  Here we find the linear trend holds up until about 1990, after which the observed temperature increases without relation to the magnetic activity.  We note that our results are in accord with other investigators~\citep{lean-3069,prsa:1452085,echera-41} and contrast with those who have found negligible evidence of solar forcing on Earth's climate~\citep{moore-L17705,moore-34918,li-1465111}.

\begin{table}[t]
\caption{Linear Correlation with CET}
\label{tab:A}
\begin{tabular}{|c|c|c|c|c|c|} \hline SMA & $N$ & $T_b$ & $\chi^2$ & $r_P$ & $\log_{10} Q$ \\\hline $R_g$ & 337 & 8.92 & 117 & 0.350 & -9.86 \\ $R_i$ & 309 & 8.85 & 107 & 0.329 & -8.15 \\ $\sqrt{R_g}$ & 337 & 8.78 & 116 & 0.358 & -10.3 \\ $\sqrt{R_i}$ & 309 & 8.54 & 110 & 0.301 & -6.92 \\\hline
\end{tabular}
\end{table}
%   WITH SQUARE ROOT
% Delta Rgpred
% ans =      0.81673
% Delta Ripred
% ans =      0.69094
% Polyval Rg
% pfit =    0.0054849        8.777
% chisqrg =        115.8
% N =   337
% nvarrg =       132.83
% pearrg =      0.35801
% Qrg =    4.96e-011
% Polyval Ri
% pfit =    0.0074024       8.5438
% chisqri =       109.52
% N =   309
% nvarri =       120.43
% pearri =      0.30103
% Qri =  1.2121e-007
% 
%   WITHOUT SQUARE ROOT (NOTE NORMALIZATION)
% Delta Rgpred
% ans =      0.82545
% Delta Ripred
% ans =      0.70517
% Polyval Rg
% pfit =     0.048002       8.9201
% chisqrg =        116.6
% N =   337
% nvarrg =       132.83
% pearrg =      0.34958
% Qrg =  1.3863e-010
% Polyval Ri
% pfit =     0.052762        8.852
% chisqri =       107.37
% N =   309
% nvarri =       120.43
% pearri =      0.32931
% Qri =  7.0885e-009
% 

\section{Discussion and Conclusions}
Many factors can influence global climate and local meteorology~\citep{barnston-1295,piscaron-1661,hameedlee-32,weber-917,lihua-231,echer-489}, thus we should not expect a correlation with $r_P = \pm 1$.  The low value of $Q$ indicates that the null hypothesis of no correlation should be rejected in favor of the linear trend.  Comparison to others' work is difficult as few authors make use of the derectified sunspot signal and most employ a cross coherence analysis of the spectra.  The methodology here is quite different, eliciting a relationship between total solar magnetic activity and the observed temperature in central England.  Naturally, investigation of the magnetic activity's correlation with other indicators of climate is forthcoming.  The position of the most recent data point in Figure~\ref{fig:L}(f) indicates a possible end to the recent warming pattern {\cbr displaying the most extreme excursion from the trend}; only time can reveal whether our energy hungry civilization has averted a disaster beyond natural proportions.

\begin{figure}[t]%[tb]
\includegraphics[width=\columnwidth]{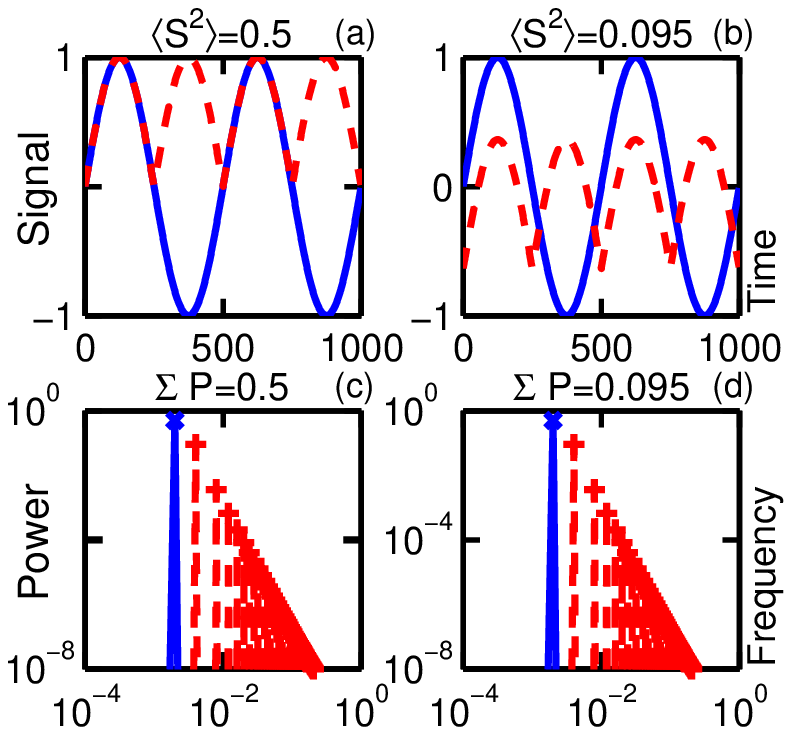}
\caption{Comparison of power spectrum of a signal ($\times$) with its rectified version ($+$) both with and without subtraction of the mean.  The power at 0 frequency equals 0.405}  %% no full stop at the end of caption
\label{fig:M}
\end{figure}

What influence has the derectification had on our results?  From comparison of the simple FFT spectrum of a sinusoidal signal with its rectified version in Figure~\ref{fig:M}, one finds that rectification turns a pure signal component into its even harmonic series and assigns most of the power to the DC component.  Consequently, spectral analysis of the absolute sunspot number will contain overtones of the Hale cycle rather than its fundamental and attribute the majority of its influence to the constant component not addressed by the wavelet transform.  The choice of where to apply the change in sign to the sunspot number generally spans only two or three years, thus introduces sporadic noise at the smallest scales which does not affect the overall result.  Ideally, one wishes to work with a sunspot number which takes polarity into account, $R_I \equiv R_+ - R_-$, where $R_\pm$ represent numbers of sunspots with opposite magnetic orientation; however, such information is not available for the time span of the historical record.

By adapting the continuous wavelet transform, one may estimate the instant power carried by a signal even as the edges are approached.  Both renormalization and window modification are found to be effective, where skewing the window improves the power estimation at the cost of frequency response and quality of reconstruction.  Analyzing both the group and international sunspot numbers produces an estimate of the level of solar magnetic activity for each.  When plotted against an ordinate of temperature, a relationship appears between the number of spots on the sun and the temperature observed in central England.  This relation holds from the middle of the 17th century up until near the end of the 20th, after which the warming trend can no longer be explained by {\cbr an increased} level of solar magnetic activity.

\begin{figure}[t]%[tb]
\includegraphics[width=\columnwidth]{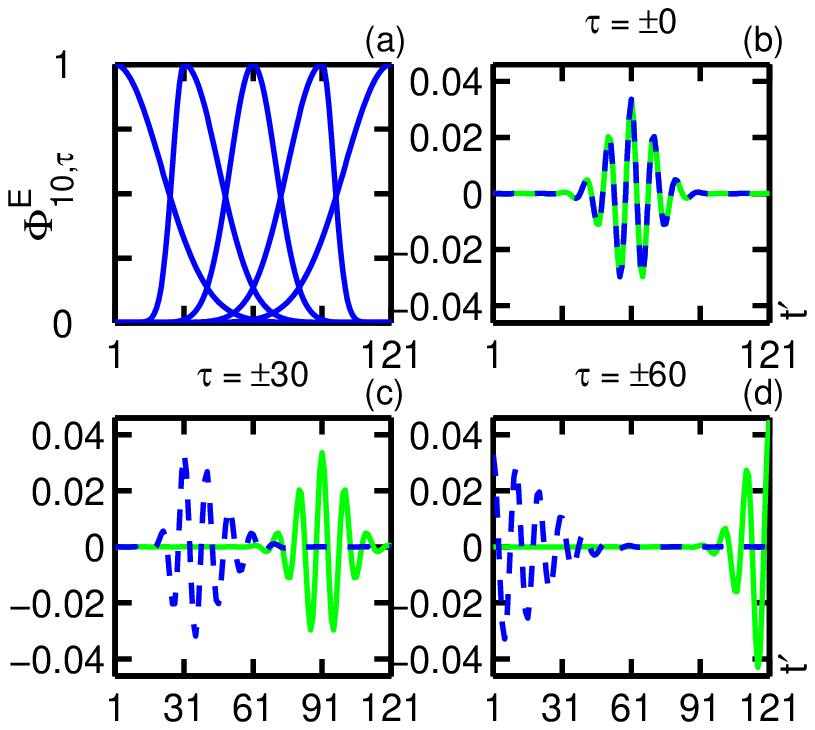}
\caption{(a) Edge adapted windows $\Phi^E$ for a scale 10 wavelet centered at the labelled ticks. The raw wavelet length is 121 units. (b)-(d) Real part of the scale 10 wavelets for the EAWT (dashed) and the RAWT (solid) for several offsets} %% no full stop at the end of caption
\label{fig:D}
\end{figure}

\begin{figure}[t]%[tb]
\includegraphics[width=\columnwidth]{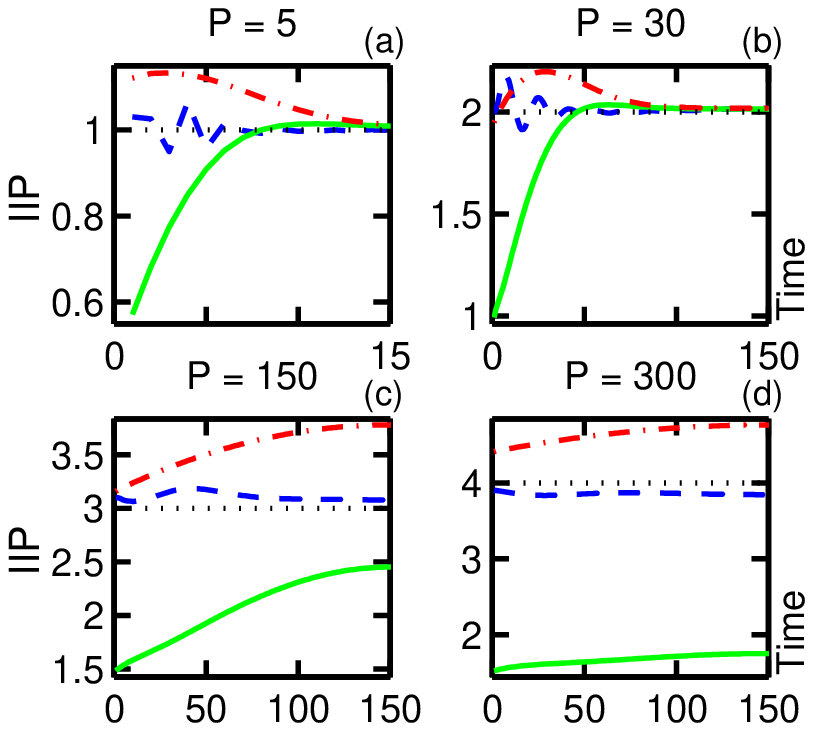}
\caption{Integrated instant power for the EAWT (dashed), the AAWT (solid), and the CCWT (dash-dot) of a test signal $y_P = Re (e^{i t 2 \pi / P})$ with period P} %% no full stop at the end of caption
\label{fig:E}
\end{figure}

\section{Appendix} 
\subsection{Renormalization and Edge Adaptation}
Compensating for the effects of signal truncation is not straightforward.  The CWT and AWT simply zero pad the signal beyond its known values, leading to the reduction in response at the edges.  Previously~\citep{rwj-jgr02}, we have suggested an algorithm to correct for the loss of response by renormalizing the CWT outside the cone-of-admissibility, and we describe it now in terms of a renormalization of the wavelet amplitude.  For wavelets truncated by either edge of the signal, the window is shifted by an offset $\tau$ relative to an unshifted window $\Phi_0$ defining the time span $t'$, and the corrected continuous wavelet transform CCWT is defined by $C_{s,t} = C_s \sqrt{\sum \Phi_0 / \sum \Phi_\tau}$, where $\sum \Phi_0 = s \sqrt{2 \pi}$ for $\sum \equiv \sum_{t'}$ when written without subscript, and similarly for wavelets truncated on both sides.  We have found lately that this algorithm may overestimate the power at low frequencies.

As~\cite{Frick:1997426} point out, wavelet truncation affects the zero mean formula, such that the continuous admissibility term is no longer adequate to remove the DC component.  The adaptive admissible wavelet transform AAWT is defined by $\Theta = \Theta^0 - d_{s,t}$, where $d_{s,t} \equiv \langle \Theta^0_\tau \rangle = \sum \Theta^0 \Phi_\tau / \sum \Phi_\tau$ is the weighted mean of the remaining wave $\Theta^0_\tau$.  Addressing now the amplitude of the truncated wavelet $\psi_\tau = C_s \Phi_\tau \Theta$, with $C_{s,t} = C_s \sqrt{2 / s^2 \vert \psi_\tau \vert^2}$ we define the renormalized adaptive wavelet transform RAWT.  These transformations are very similar to those encountered when building operators in lattice gauge theory~\citep{johnson:074502} and in other applications.

Finally, we consider an algorithm which modifies the shape of the analyzing window $\Phi$ with respect to the location and length of the signal.  The length of a truncated wavelet $\psi_\tau$ is defined to be the lesser of the raw wavelet length or the signal length, $N_\tau = \min (N_{t'},N_t)$.  The offset $\tau(t)$ is determined from either the center of the signal or the location of the cone-of-admissibility, and the algorithm to keep everything aligned gets a bit complicated: for $\tau' = \max (0 , \lfloor s \chi - \lfloor N_t/2 )$ and $t'(k) = [- \lfloor s \chi, \lfloor s \chi]$ indexed by $k$ from 1 to $N_{t'}$, if $\tau \leq 0$ then $t' \rightarrow t'[1, \min (N_{t'},N_t)] + \tau'$, else $t' \rightarrow t'[\max(1,N_{t'}-N_t+1), N_{t'}] - \tau'$.  The end result is simply to truncate either edge of the wavelet as necessary.  Then, for $\gamma = \min (\lfloor s \chi , \lfloor N_t/2)$, the edge adapted window is $\ln \Phi^E_{s,t}(t') = -[\eta \gamma / (\gamma \pm \tau)]^2/2$ as $t' <$ or $> \tau$, respectively, where $\Phi^E_{s,t}(\tau) = 1$, which scales the $t'$ axis of the window to either side of $\tau$.  These wavelets, renormalized as above, define the edge adapted wavelet transform EAWT.  In Figure~\ref{fig:D}(a) we show the adapted windows for a scale 10 wavelet centered at the labelled ticks, and in Figures~\ref{fig:D}(b)-(d) we display the real part of that wavelet for the EAWT and the RAWT for several offsets.

\begin{figure}[t]%[tb]
\includegraphics[width=\columnwidth]{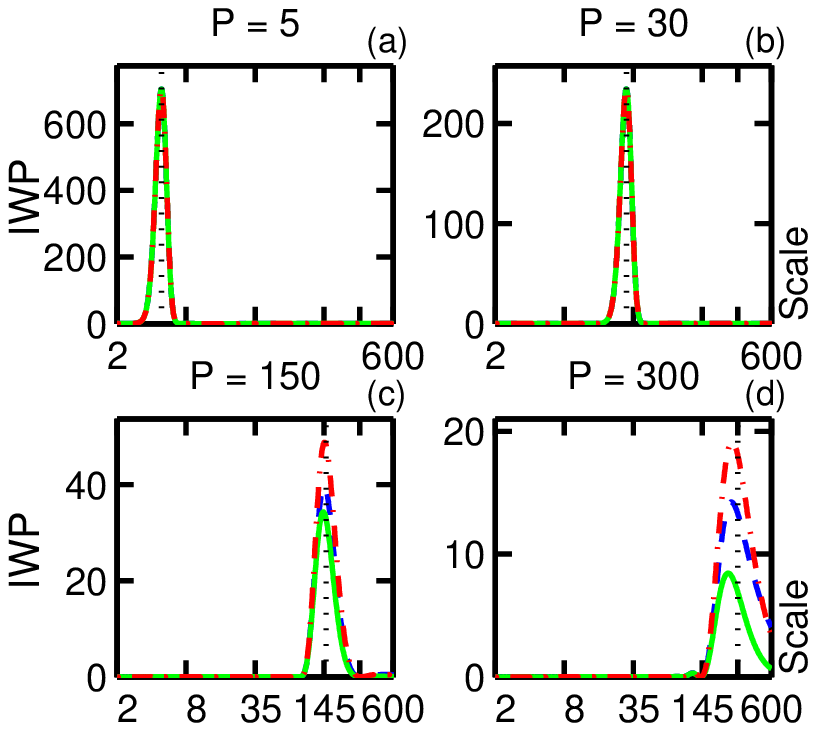}
\caption{Instant wavelet power at $N_t/2$ for test signals with period P indicated by the vertical dotted line for the EAWT (dashed), the AAWT (solid), and the CCWT (dash-dot).  Note that the y-axes are in milliunits.  In succession, the integrated power for the EAWT is 1.01, 2.04, 3.08, and 3.84} %% no full stop at the end of caption
\label{fig:F}
\end{figure}
% Fig F -- IIP for E A N:
% iipfnt2 =
%         1.005        1.005       1.0051
%        2.0374        2.029       2.0492
%        3.0789       2.4522       3.7795
%        3.8448       1.7523        4.763

\begin{figure}[t]%[tb]
\includegraphics[width=\columnwidth]{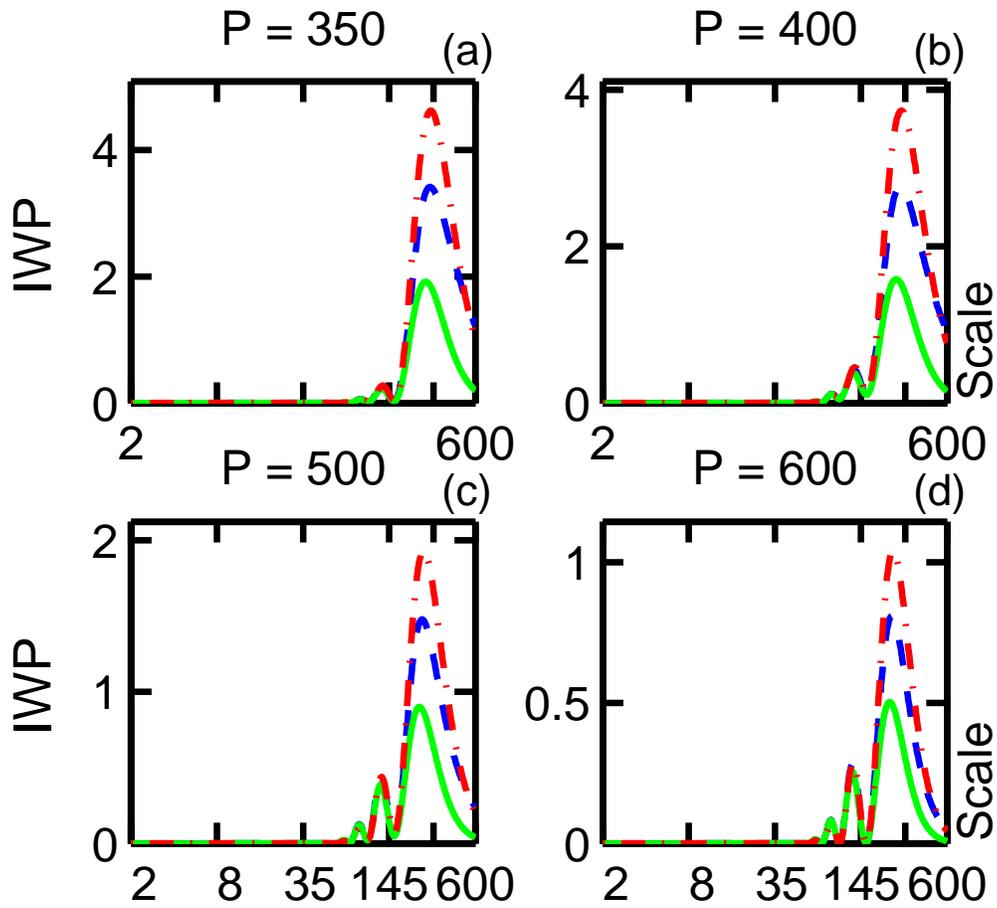}
\caption{Instant wavelet power at the midpoint of the duration for test signals with period P for the EAWT (dashed), the AAWT (solid), and the CCWT (dash-dot).  Note that the y-axes are in milliunits.  In succession, the integrated power for the EAWT is 0.97, 0.78, 0.34, and 0.15} %% no full stop at the end of caption
\label{fig:G}
\end{figure}
% Fig G -- IIP for E A N:
% iipfnt2 =
%       0.97304      0.42047         1.19
%       0.78275      0.34933      0.95629
%       0.34188      0.17094      0.42286
%       0.15395     0.083181      0.19296

\begin{table}[t]
\caption{Summary of the wavelet transforms}
\label{tab:B}
\begin{tabular}{|c|c|c|c|} \hline $\psi = C \Phi \Theta$ & $C_{s,t}$ & $\Phi_{s,t}(t')$ & $\Theta_{s,t}(t')$ \\\hline
% CWT & $C_s = \sqrt{2}\, C^0_s / s$ & $\Phi^0 = e^{- \eta^2/2}$ & $\Theta^0 = e^{i \omega_1 \eta}$ \\
 CWT & $C_s$ & $\Phi^0$ & $\Theta^0$ \\
 AWT & $C_s$ & $\Phi^0$ & $\Theta^0 - d_c$ \\
 CCWT & $C_s \sqrt{\sum \Phi_0 / \sum \Phi_\tau}$ & $\Phi^0$ & $\Theta^0$ \\
 AAWT & $C_s$ & $\Phi^0$ & $\Theta^0 - d_{s,t}$ \\
 RAWT & $C_s \sqrt{2 / s^2 \vert \psi_\tau \vert^2}$ & $\Phi^0$ & $\Theta^0 - d_{s,t}$ \\
 EAWT & $C_s \sqrt{2 / s^2 \vert \psi_\tau \vert^2}$ & $\Phi^E$ & $\Theta^0 - d_{s,t}$ \\\hline
\end{tabular}
\end{table}

\subsection{Comparison of the Adapted Transforms}
We next embark upon a lengthy exploration of the relative responses of the various transforms.  The properties of our transform menagerie are summarized in Table~\ref{tab:B}.  Let us first consider the responses of the EAWT, the AAWT, and the CCWT to a succession of test signals $y_k$ of duration $N_t = 300$ with squared amplitudes $A_k^2 = [1,\, 2,\, 3,\, 4]$ and periods $P_k = [5,\, 30,\, 150,\, 300]$.  In Figure~\ref{fig:E} we display the IIP for each test signal.  Generally, the CCWT is found to overcompensate for the wavelet truncation while the EAWT displays some oscillation in the IIP near the edge.  A slight overestimation of the power is attributed to the signal truncation.  The IIP for the AAWT drops significantly in response to the wavelet truncation, especially for signal periods wholly beyond the cone-of-influence, while the IIP for the EAWT remains fairly close to the signal power for the duration.  In Figure~\ref{fig:F} we show the IWP at the midpoint of the transform for each of the test signals.  Here, the peaks differentiate as the signal period approaches the signal length.  We notice in Figure~\ref{fig:F}(d) that the peaks are not located precisely at the signal period.  Intrigued, we next consider a series of test signals of unit amplitude with periods $P_k = [350,\, 400,\, 500,\, 600]$ exceeding the duration, Figure~\ref{fig:G}.  As the signal period increases, the magnitude of the response diminishes, and the location of the peaks seems bound from above by the scale of the duration $N_t$.  Additional peaks appear at harmonics of the response for these extremely low frequency ELF signals.  Such behavior is not unexpected in light of Figure~\ref{fig:C}, where one sees that the wavelet response to ELF signals has a finite bound in scale.

\begin{figure}[t]%[tb]
\includegraphics[width=\columnwidth]{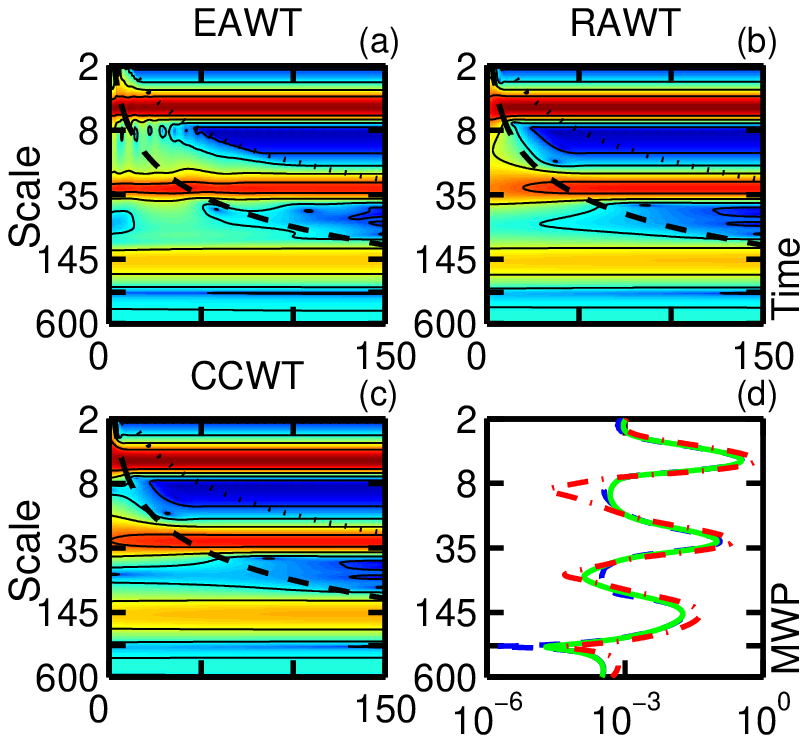}
\caption{{\cbr Power spectral density (a)-(c) and  mean wavelet power (d) of the test signal $y = y_5 + y_{30} + y_{150}$ for the EAWT (dashed), the RAWT (solid), and the CCWT (dash-dot)} } %% no full stop at the end of caption
\label{fig:H}
\end{figure}

\begin{figure}[t]%[tb]
\includegraphics[width=\columnwidth]{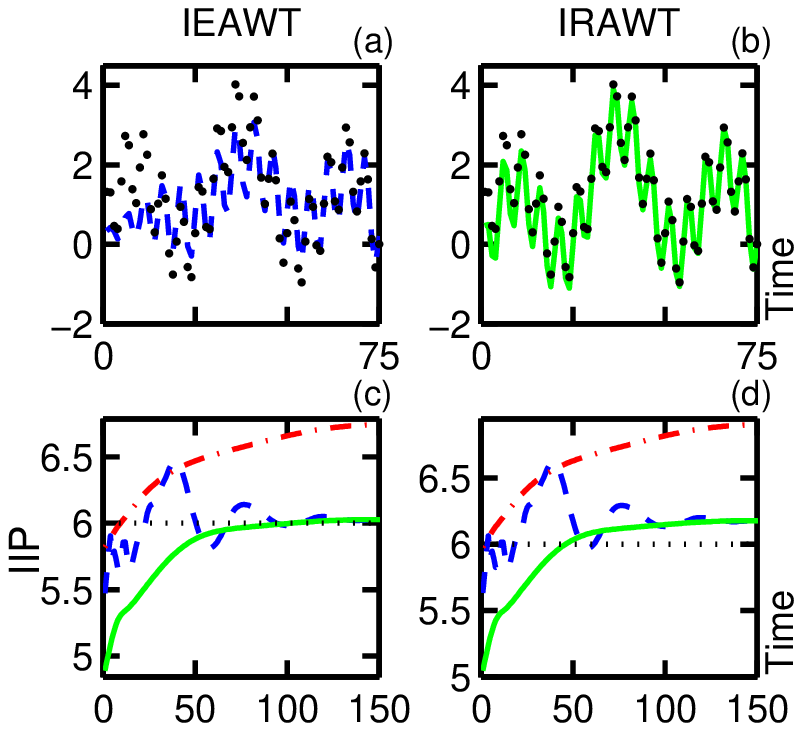}
\caption{{\cbr Reconstruction from the IEAWT (a) and the IRAWT (b) and integrated instant power (c)-(d) for the EAWT (dashed), the RAWT (solid), and the CCWT (dash-dot).  The IIP of (c) has been cutoff at the scale $N_t = 300$, while that of (d) includes the ELF contribution} } %% no full stop at the end of caption
\label{fig:N}
\end{figure}

Finally, we consider the responses of the EAWT, the RAWT, and the CCWT to a test signal of the same duration comprised of the sum $y = \sum_{k=1}^3 y_k$ of the first three signals above, with their PSD displayed in Figures~\ref{fig:H}(a)-(c).  The CCWT produces a deeper trough between the peaks in the MWP of Figure~\ref{fig:H}(d) and a generally stronger response, while the EAWT returns a deeper trough at the scale of the duration.  In Figures~\ref{fig:N}(a) and (b) we display the reconstructions for the newly proposed transforms.  The loss of reconstruction for the IEAWT is of a different character than for the others, and we wonder if an alternate reconstruction basis would be more effective.  The best reconstruction is provided by the IRAWT, which still displays some edge alterations.  The IIPs of Figure~\ref{fig:N}(c) have been evaluated with a cutoff at $N_t$, while those of (d) include the contribution from the ELF region.  The effect of the renormalization on the RAWT is clearly seen, as now its IIP agrees with that of the EAWT at the midpoint $N_t/2$, while the CCWT overestimates the signal power.  The ELF contribution is apparent as an increase in the IIP above the instant signal power and is attributable to the effect of signal truncation.  We conclude that the EAWT and the RAWT each provide a reasonable estimate of the instant signal power.

%%% END ARTICLE TEXT

%\nolinenumbers

%% Acknowledgements
% \acknowledgments
\subsection{Acknowledgments}
International sunspot number provided by the SIDC-team, World Data Center for the Sunspot Index, Royal Observatory of Belgium, Monthly Report on the International Sunspot Number, online catalogue of the sunspot index: http://www.sidc.be/sunspot-data/, 1700--2008.  Group sunspot number provided by Solar Data Services, National Geophysical Data Center, National Oceanic and Atmospheric Administration, USA.  Global temperature data provided by J. Oerlemans, 2005, Global Glacier Length Temperature Reconstruction, IGBP PAGES/World Data Center for Paleoclimatology, Data Contribution Series \#2005-059, NOAA/NCDC Paleoclimatology Program, Boulder CO, USA.  Central England Temperature record provided by the Met Office Hadley Centre for Climate Change, http://www.metoffice.gov.uk/hadobs.

%% References
%% Please cite all reference entries in the article text using \cite or
%% equivalent command. 
%%%  Using BibTeX  (Name-Year style)
%
%\bibliographystyle{spr-mp-nameyear-cnd}  %% BibTeX style
%\bibliographystyle{unsrt}
%\bibliography{../solar,../particle}                %% BibTeX data

%\newpage
%\listoffigures

\end{document}